\documentstyle[amssymb]{article}

\begin{document}

\begin{center}

{\bf \Large The Variable Coefficient Hele-Shaw Problem, Integrability and Quadrature Identities}\\

\vspace{5mm}

{\Large Igor Loutsenko}\\

\vspace{3mm}

Institute of Mathematics, University of Oxford, 24-29 st. Gilles',
Oxford, OX1 3LB, UK\\

\vspace{5mm}

e-mail loutsenk@maths.ox.ac.uk

\vspace{5mm}

Abstract

\end{center}

\begin{quote}

The theory of quadrature domains for harmonic functions and the
Hele-Shaw problem of the fluid dynamics are related subjects of the
complex variables and mathematical physics. We present results
generalizing the above subjects for elliptic PDEs with variable
coefficients, emerging in a class of the free-boundary problems for
viscous flows in non-homogeneous media. Such flows posses an
infinite number of conservation laws, whose special cases may be
viewed as quadrature identities for solutions of
variable-coefficient elliptic PDEs. If such PDEs are gauge
equivalent to the Laplace equation (gauge-trivial case), a
time-dependent conformal map technique, employed for description of
the quadrature domains, leads to differential equations, known as
``string" constraints in the theory of integrable systems. Although
analogs of the string constraints have non-local forms for
gauge-non-trivial equations, it is still possible to construct the
quadrature domains explicitly, if the elliptic operator belongs to a
class of the Calogero-Moser Hamiltonians.

\end{quote}

\begin{section}{Introduction}

Recently, the constant-coefficient Hele-Shaw problem has received a
good deal of attention due to its connection with the theory of
integrable hierarchies in the dispersionless limit \cite{MWZ},
\cite{WZ}. These integrable structures turned out to have natural
interpretations in the classical free-boundary problems.

Although concerned with various aspects of integrability, the present
paper is primarily devoted to construction of solutions for a more general
class of the variable-coefficient
free-boundary problems. Namely, motivated by practical applications,
we derive a class of rational solutions of the variable-coefficient Hele-Shaw problems, related to
the dihedral Calogero-Moser systems. These solutions provide
examples of quadrature identities for elliptic PDEs with variable
coefficients. The theory of quadrature domains for such PDEs is
highly reminiscent of that for harmonic functions.

The structure of the paper is as follows: A brief summary of the theory of the
Hele-Shaw flows, as well as that of the quadrature domains, is
given in the next three sections, followed by introduction of the
time-dependent conformal map technique in Section 5. We digress into
consideration of the gauge-trivial problems, considered in the above
mentioned works on relationship between dispersionless integrable
hierarchies and the Hele-Shaw problem, in Section 6. Sections 7 and
8 are devoted to derivation of explicit results for a class of the
gauge-non-trivial problems. Concluding remarks are given in Section
9.

\end{section}

\begin{section}{Variable Coefficient Hele-Shaw Problem}

The Hele-Shaw flow governs the dynamics of the boundary $\partial
\Omega=\partial \Omega(t)$ in the plane separating two disjoint,
open regions $\Omega=\Omega(t)$ and ${\mathbb
C}\backslash\bar\Omega$ in which scalar fields are defined. These
may be interpreted as the pressure fields for two incompressible
(viscous and non-viscous) immiscible fluids, trapped in a gap
between two plates or propagating in a layer of the porous medium.
If the size of the gap or the layer height is negligible, the
velocity of the viscous liquid can be averaged across the direction
perpendicular to the layer or the plates respectively, and the
problem becomes effectively two dimensional. The velocity of the
viscous liquid is then proportional to the two dimensional pressure
gradient (the Darcy law)
\begin{equation}
v=-\nabla P \label{uniform}
\end{equation}
The flow is incompressible and the liquid velocity satisfies a
continuity equation
\begin{equation}
\nabla\cdot v=0 \label{con}
\end{equation}
The region $\Omega$ can be chosen to be bounded and occupied with a
viscous liquid, so that (\ref{uniform}) holds in that region. It
will be referred as the ``interior" region. It is surrounded by a
non-viscous liquid occupying the ``exterior" (unbounded) region
${\mathbb C}\backslash\bar\Omega$. Since the flow of the liquid
occupying the exterior domain is inviscid, it must be driven by the
pressure field with vanishing gradient and, therefore, the pressure
is constant in that region. Without loss of generality the pressure
can be set to zero in ${\mathbb C} \backslash \bar \Omega$. It is a
continuous function across the moving boundary $\partial\Omega(t)$,
so that
\begin{equation}
P(\partial\Omega)=0 \label{partial}
\end{equation}
and the normal velocity of the boundary coincides with that of the
viscous flow at $\partial\Omega$
\begin{equation}
n\cdot\frac{d}{dt}\partial\Omega=n\cdot v \label{velocities}
\end{equation}
where $n$ denotes the outward normal to the boundary. In the present
and the next section we consider flows that are driven by a single
point source located inside $\Omega$. Without loss of generality we
can locate it at the origin $z=0$
$$
P\to \frac{-1}{4}\log(z\bar z), \quad {\rm as} \quad z\to 0
$$
where $z:=X+iY$, and $X,Y$ stand for the Cartesian coordinates on
the plane. The last equation, together with the Darcy law
(\ref{uniform}) and the continuity equation (\ref{con}) leads to
$$
\Delta P=-\pi\delta(X)\delta(Y)
$$
where $\delta$ denotes the Dirac delta-function. In this paper we
deal with a generalization of the above problem, namely the
Hele-Shaw problem with coefficients depending spatial variables $X$
and $Y$
\begin{equation}
v=-\kappa \nabla P, \quad \nabla \cdot (\eta v)=0 \label{norm}
\end{equation}
where $\kappa=\kappa(X,Y), \eta=\eta(X,Y)$ are arbitrary functions
of $X,Y$, sufficiently regular in $\Omega$. As in the constant
coefficient case, the pressure has a logarithmic singularity at the
origin, i.e.
\begin{equation}
P\to \frac{-1}{4\kappa(0)\eta(0)}\log(z\bar z), \quad {\rm as} \quad
z\to 0 \label{origin}
\end{equation}
when the flow is driven by a single point source.

Note, that one has to modify the asymptotic condition (\ref{origin})
if $\kappa\eta$ vanishes or has a singularity at the point source,
or if several point sources coalesce in a special way (``multipole"
sources).

We consider a situation when the exterior region is occupied by a
non-viscous liquid, so that (\ref{partial}) holds. Equations
(\ref{partial}), (\ref{velocities}), (\ref{norm}), as well as
asymptotic conditions (e.g. (\ref{origin}), for a single point
source) set the Hele-Shaw problem with variable coefficients,
describing, for instance, propagation of the liquid in a thin
horizontal layer of non-homogeneous porous medium, whose
permeability and height depend on $X,Y$.

\end{section}

\begin{section}{Conservation Laws}

From (\ref{norm}), (\ref{origin}) it follows that the pressure
satisfies the second order elliptic PDE
\begin{equation}
(\nabla \kappa\eta\nabla)P=-\pi\delta(X)\delta(Y) \label{nabla}
\end{equation}
that replaces the Laplace equation of the constant coefficient
problem. Let $\phi(z,\bar z)$ be any regular in the interior domain
solution of the equation
\begin{equation}
(\nabla \kappa \eta\nabla)\phi=\left(\frac{\partial}{\partial
z}\kappa\eta\frac{\partial}{\partial \bar
z}+\frac{\partial}{\partial \bar
z}\kappa\eta\frac{\partial}{\partial z}\right)\phi=0
\label{schrodinger}
\end{equation}
Introduce the following quantities
\begin{equation}
M[\phi]=\int_{\Omega(t)}\eta\phi dXdY \label{integrals}
\end{equation}
and estimate their time derivatives. Considering an infinitesimal
variation of the interior domain $\Omega(t)\to \Omega(t+dt)$ along
the boundary $\partial\Omega(t)$, we get
$$
\frac{d M[\phi]}{dt}=\oint_{\partial \Omega(t)}\eta\phi v_ndl
$$
where $dl$ is the boundary length element and $v_n$ is the normal
velocity of the boundary. From (\ref{velocities}) and (\ref{norm})
it follows that $v_n=v\cdot n=-\kappa n\cdot\nabla P$. Since
$P(\partial \Omega)=0$
$$
\frac{d M[\phi]}{dt}=\oint_{\partial
\Omega(t)}\left(P\kappa\eta\nabla\phi-\phi\kappa\eta\nabla
P\right)\cdot ndl
$$
Applying the Stokes theorem and taking (\ref{nabla}) and
(\ref{schrodinger}) into account, from the last equation, we get
$$
\frac{dM[\phi]}{dt}=\oint_{z=0}\left(P\kappa\eta\nabla\phi-\phi\kappa\eta\nabla
P\right)\cdot ndl
$$
that, according to (\ref{origin}), becomes
\begin{equation}
\frac{dM[\phi]}{dt}=\pi\phi(0) \label{linear}
\end{equation}
Therefore, the quantity $M[\phi]$ is conserved for any (regular in
$\Omega$) solution of (\ref{schrodinger}), such that $\phi(0)=0$.

The Richardson harmonic moments $\int_{\Omega(t)} z^kdXdY$ \cite{R}
of the constant coefficient Hele-Shaw problem correspond to the
special case
$$
\kappa=\eta=1, \quad \phi(z,\bar z)=z^k, \quad k\ge 0
$$
To the author knowledge, the conservation laws for the
variable-coefficient problem were first presented in \cite{EE}.

\end{section}

\begin{section}{Quadrature Domains}

The derivation of conservation laws can be easily generalized to the
case of several sources. Consider the flow driven by $N$ sources of
the time-dependent powers $q_k(t)$ that are located at points $z_k,
k=1..N$ in $\Omega$
$$
P\to \frac{-q_k(t)} {2\kappa(x_k,y_k)\eta(x_k,y_k)}\log|z-z_k|,
\quad {\rm as} \quad z\to z_k, \quad z_k=x_k+iy_k\in \Omega
$$
Then by arguments, similar to those used for the single point source
problem we arrive at
$$
\frac{dM[\phi]}{dt}=\pi\sum_{k=1}^N q_k(t)\phi(z_k,\bar z_k)
$$
for any $\phi(z,\bar z)$, regular in $\Omega$ and satisfying
(\ref{schrodinger}). It follows that
$$
M[\phi](t)=M[\phi](0)+\pi\sum_{k=1}^N Q_k\phi(z_k,\bar z_k), \quad
Q_k=Q_k(t)=\int_{0}^{t} q_k(t')dt'
$$
From the last equation, we see that $M[\phi](t)$ does not depend on
history of sources and is a function of total charges $Q_i$,
produced by the time $t$. This fact reflects the integrability of
the problem, where flows produced by different sources commute.

Consider the special case when $M[\phi](0)=0$. It describes the
injection of the fluid to an initially empty Hele-Shaw cell. Taking
into account the definition (\ref{integrals}) of $M[\phi]$ we obtain
\begin{equation}
\int_\Omega\eta\phi(z,\bar z)dXdY=\pi\sum_{k=1}^N Q_k\phi(z_k,\bar
z_k) \label{identity}
\end{equation}
This is an identity expressing the integral over the domain $\Omega$
in the left-hand side as a sum of terms evaluated at a finite number
of points, given on the right-hand side. The special case
$\eta=\kappa=1$ provides quadrature identities for harmonic
functions. Special domains $\Omega$, possessing the above property,
are called quadrature domains \cite{Sa}, \cite{Sh}, \cite{VE}. The simplest
example of a quadrature domain is a circular disc, produced by a
single point source in the constant coefficient Hele-Shaw problem.
The corresponding quadrature identity is a ``mean value'' theorem
for harmonic functions. Equation (\ref{identity}) is a
generalization of quadrature identities appearing in the theory of
harmonic functions to the case of elliptic equations with variable
coefficients. The quadrature domains for such PDEs are, thus,
solutions to the variable coefficient, interior Hele-Shaw problems
with zero initial conditions. In the sequel we will mainly deal with
situation when groups of sources coalesce in such a way that
(\ref{identity}) becomes
\begin{equation}
\int_\Omega\eta\phi(z,\bar z)dXdY=\pi\sum_{k=1}^N\hat
Q_k[\phi](z_k,\bar z_k) \label{identity1}
\end{equation}
Where ${\hat Q}_k$ is a finite-order, differential in $z, \bar z$
operator of the following form
\begin{equation}
{\hat
Q}_k=Q^{(0)}_k+\sum_{i=1}^{i_k}\left(Q_k^{(i)}\frac{\partial^i}{\partial
z^i}+\bar Q_k^{(i)}\frac{\partial^i}{\partial \bar z^i}\right),\quad
Q_k^{(0)}=\bar Q_k^{(0)}, \label{Quad}
\end{equation}
For such combinations of multipole sources, the integral over
$\Omega$ in the left-hand side of (\ref{identity1}) is expressed as
a sum of terms involving values of function $\phi$ as well as a
finite number of its derivatives at a finite number of points inside
$\Omega$. The boundary $\partial \Omega(t)$ of the quadrature domain
is a solution to the variable coefficient Hele-Shaw problem with
pressure satisfying the following equation
$$
\nabla\kappa\eta\nabla P=-\pi\sum_{k=1}^N\hat
q_k[\delta(X-x_i)\delta(Y-y_i)], \quad \hat q_k=\frac{\partial
Q^{(0)}_k}{\partial t}+\sum_{i=1}^{i_k}(-1)^i\left(\frac{\partial
Q^{(i)}_k}{\partial t}\frac{\partial^i}{\partial z^i}+\frac{\partial
\bar Q_k^{(i)}}{\partial t}\frac{\partial^i}{\partial \bar
z^i}\right)
$$
Note, that operators $\hat Q_k$ do not contain mixed derivatives
$\frac{\partial^{n+m}}{\partial \bar z^n\partial z^m}$, since, due
to (\ref{schrodinger}), $\frac{\partial^2\phi}{\partial z\partial
\bar z}$ is a linear combination of the first order derivatives of
$\phi$. The coefficient in front of $\frac{\partial^i}{\partial
z^i}$ must be complex conjugate of that in front of
$\frac{\partial^i}{\partial \bar z^i}$, since both $\phi(z,\bar z)$
and its complex conjugate satisfy $(\ref{schrodinger})$.

\end{section}

\begin{section}{Time-Dependent Conformal Maps}

Usually, a time dependent conformal map technique is implemented to
find explicit solutions to the constant coefficient Hele-Shaw
problem and, in the special case of zero initial conditions, it is
also an efficient method of constructing quadrature domains. This
technique is also applicable to the the variable coefficient case,
leading to explicit solutions of some non-trivial problems.

Introduce a ``mathematical" $w$-plane and denote by $z(w,t)$ a
conformal map from the unit disc $|w|<1$ in the $w$-plane to a
simply-connected interior region $\Omega(t)$ in the physical $z$
plane. According to the Riemann mapping theorem a one-to one
analytic in $|w|\le 1$ map
\begin{equation}
z(w,t)=r(t)w+\sum_{i>0}u_i(t)w^{i+1} \label{transform}
\end{equation}
exists, such that the unit circle in the $w$-plane is mapped to the
(analytic) boundary contour in the $z$-plane
\begin{equation}
z(w,t) \in \partial\Omega(t), \quad {\rm if} \quad |w|=1
\label{maptoboundary}
\end{equation}
Alternatively to the derivation of Section 2, one can estimate the
time derivatives of $M[\phi]$, transforming the two-dimensional
integrals (\ref{integrals}), taken over $\Omega$, to line integrals
along the unit circle in the $w$-plane. Introducing a function
$\xi(z,\bar z)$ such that
\begin{equation}
\eta(X,Y)\phi(z,\bar z)=\frac{\partial}{\partial \bar z}\xi(z,\bar
z) \label{eint}
\end{equation}
by the Green theorem and (\ref{maptoboundary}), we may rewrite
(\ref{integrals}) as
\begin{equation}
M[\phi]=\frac{1}{2i}\oint_{\partial\Omega}\xi
dz=\frac{1}{2i}\oint_{|w|=1}\xi\frac{\partial z}{\partial w}dw
\label{line}
\end{equation}
Note that $r(t)$ in (\ref{transform}) can be made to be real and
\begin{equation}
\bar w=1/w, \quad \bar z=\bar
z(1/w,t)=\frac{r(t)}{w}+\sum_{i>0}\frac{\bar u_i(t)}{w^{i+1}}
\label{transform1}
\end{equation}
along the boundary. In the sequel we mainly deal with $z(w,t)$
evaluated at the boundary $\bar w=1/w$ and therefore we use $\bar z$
to denote $\bar z(1/w,t)$ (if not otherwise specified). It follows
from (\ref{eint}) and (\ref{line}) that
$$
\frac{dM[\phi]}{dt}=\frac{1}{2i}\oint_{|w|=1}\eta\phi\left(\frac{\partial
z}{\partial w}\frac{\partial \bar z}{\partial t}-\frac{\partial
z}{\partial t}\frac{\partial \bar z}{\partial
w}\right)dw+\frac{1}{2i}\oint_{|w|=1}\frac{\partial}{\partial
w}\left(\xi\frac{\partial z}{\partial t}\right)dw
$$
which equals
\begin{equation}
\frac{dM[\phi]}{dt}=\frac{1}{2i}\oint_{|w|=1}\phi\eta\frac{\{z(w,t),\bar
z(1/w,t)\}}{w}{dw} \label{evolution}
\end{equation}
provided $\xi$ in (\ref{eint}) is univalent at $\partial\Omega$. In
(\ref{evolution}), $\{,\}$ denotes the Poisson bracket
\begin{equation}
\{f(w,t),g(w,t)\}:=w\frac{\partial f}{\partial w}\frac{\partial
g}{\partial t}-w\frac{\partial g}{\partial w}\frac{\partial
f}{\partial t} \label{lax bra}
\end{equation}
defined on the cylinder parameterized by coordinates
$(w=e^{i\theta},t)$ with real $\theta$ and $t$. By virtue of
(\ref{evolution}) the evolution equation (\ref{linear}) may be
rewritten as
\begin{equation}
\frac{1}{2\pi i}\oint_{|w|=1}\phi\eta\frac{\{z,\bar
z\}}{w}dw=\phi(0) \label{linearevolution}
\end{equation}

\end{section}

\begin{section}{Gauge Trivial Problems, String Equations}

Consider equation (\ref{schrodinger}). Its solution $\phi$ also
satisfies
\begin{equation}
L[\phi]=0, \quad L:=\psi^{-2}\frac{\partial}{\partial
z}\psi^2\frac{\partial}{\partial \bar
z}+\psi^{-2}\frac{\partial}{\partial \bar
z}\psi^2\frac{\partial}{\partial \bar z}, \quad
\psi:=\sqrt{\kappa\eta}, \quad \psi(z, \bar z)=\bar \psi(\bar z, z)
\label{def}
\end{equation}
The elliptic second order differential operator $L$ is amenable, by
a gauge transformation, to the two-dimensional zero-magnetic field
Schroedinger operator
\begin{equation}
H:=-\psi L\frac{1}{\psi}=-\frac{1}{2}\Delta+V(X,Y), \quad
V=\frac{\Delta \psi}{2\psi}, \quad H[\psi\phi]=0, \quad {\rm if}
\quad L[\phi]=0 \label{operator}
\end{equation}
The potential $V$ vanishes when $\psi=h$, where $h$ stands for a
harmonic function
\begin{equation}
L={\mathcal L}_0:=h^{-1}\Delta h, \quad {\mathcal
L}_0\left[\frac{z^k}{h}\right]=0, \quad k\ge 0 , \quad
h:=\Psi(z)+\bar\Psi(\bar z) \label{use}
\end{equation}
The elliptic operator ${\mathcal L}_0$ is gauge equivalent to the
Laplace operators. We call such operators and corresponding
Hele-Shaw problems ``gauge-trivial". According to (\ref{integrals}),
(\ref{use}) the following quantities
$$
I_0=\int_{\Omega(t)}\frac{\eta}{h} dX dY, \quad
I_k=\int_{\Omega(t)}\frac{\eta}{h} z^k dX dY, \quad k>0
$$
are linear and constant in time.

Using (\ref{use}), from (\ref{linearevolution}) we see, that in the
gauge-trivial case
$$
\oint_{|w|=1}\frac{z^k}{w}\frac{\eta}{h}\{z,\bar
z\}dw=\oint_{|w|=1}\frac{\bar z^k}{w}\frac{\eta}{h}\{z,\bar
z\}dw=2\pi i\delta_{k0}, \quad k\ge 0
$$
or, taking (\ref{transform}), (\ref{transform1}) into account
$$
\oint_{|w|=1}\left(\frac{\eta}{h}\{z,\bar
z\}-1\right)\frac{z^k}{w}dw=\oint_{|w|=1}\left(\frac{\eta}{h}\{z,\bar
z\}-1\right)\frac{\bar z^k}{w}dw=0
$$
The expression in the parentheses of the last equation must vanish,
since $z^k/w=r^kw^{k-1}+...$ and $\bar z^k=r^kw^{-k-1}+...$, $k\ge
0$ form a basis of an arbitrary Laurent series in $w$. Therefore,
$z, \bar z$ satisfy the following differential equation
$$
\{z(w,t),\bar z(1/w,t)\}=h/\eta
$$
or equivalently
\begin{equation}
\{q(z,\bar z),\bar q(\bar z,z)\}=1, \quad (\partial_{z}
q)(\partial_{\bar z} \bar q)-(\partial_z\bar q)(\partial_{\bar z}
q)=h/\eta \label{string}
\end{equation}
known as the ``string" constraint in the theory of the
dispersionless integrable hierarchies \cite{HLY}, \cite{MWZ},
\cite{TT}, \cite{WZ}. In the special case $h=1, \eta=1$, or
equivalently $q(z,\bar z)=z$, corresponding to the constant
coefficient Hele-Shaw problem, the conformal maps satisfy the
following constraint
$$
\{z(w,t),\bar z(1/w,t)\}=1
$$
known as a Galin-Polubarinova equation \cite{G}, \cite{P} in the
theory of the Hele-Shaw flows.

The string equation (\ref{string}) is preserved by the Lax-Hamilton
flows
$$
\frac{dz}{d\tau}=\{z,{\mathcal H}\}, \quad \frac{d\bar
z}{d\tau}=\{\bar z,{\mathcal H}\}, \quad {\mathcal H}={\mathcal
H}(w,t,\tau)
$$
Indeed,
$$
\frac{d}{d\tau}\{q(z,\bar z),\bar q(\bar z,
z)\}=\left\{\frac{dq(z,\bar z)}{d\tau},\bar q(\bar
z,z)\right\}+\left\{q(z,\bar z),\frac{d\bar q(\bar
z,z)}{d\tau}\right\}=\{{\mathcal H},\{q(z, \bar z), \bar q(\bar z,
z)\}\}=0
$$
i.e. functions $z(w,t), \bar z(1/w,t)$, satisfying (\ref{string})
belong to invariant, under the action of the Lax-Hamilton vector
fields $\{{\mathcal H},\cdot\}$, subspace of space of functions of
$w,t$. It is also necessary to satisfy the condition of the
form-invariance of $z(w,t)$ along the Lax-Hamilton flow lines, i.e.
such Lax-Hamilton functions must be chosen, that $z(w,t)$ will
remain the Taylor series (\ref{transform}) along the corresponding
flows. So selected Lax-Hamilton functions, generate symmetry
transformations of the Hele-Shaw problem, mapping continuously one
solution of the problem into the others. An ableian subset of these
transformations, forms a parametrizable set of deformations, leaving
ivariant equations governing the Hele-Shaw flow. The Lax-Hamilton
functions generating such abelian subset can be conveniently chosen
as
$$
H_k=(z^{-k})_{<0}+1/2(z^{-k})_0, \quad \bar H_k=(\bar
z^{-k})_{>0}+1/2(\bar z^{-k})_0, \quad k>0
$$
where $()_{>0}, ()_{<0}, ()_0$ stand for negative, positive and zero
parts of the Laurent expansions
$$
(f)_{>0}:=\sum_{k>0}f_kw^k, \quad (f)_0:=f_0, \quad
(f)_{<0}:=\sum_{k<0}f_kw^k, \quad {\rm if} \quad
f=\sum_{k\in{\mathbb Z}}f_kw^k
$$
The flow equations
\begin{equation}
\begin{array}{cc}
\frac{\partial z}{\partial \tau_k} = \{{\mathcal H}_k , z\} & \qquad \frac{\partial z}{\partial\bar \tau_k}  = \{\bar{\mathcal H}_k , z\}  \\
\frac{\partial \bar z}{\partial \tau_k} = \{{\mathcal H}_k ,
\bar{z}\} & \qquad \frac{\partial\bar z}{\partial \bar \tau_k} =
\{\bar{\mathcal H}_k , \bar z\}
\end{array}, \quad k>0
\label{toda}
\end{equation}
constitute the two-dimensional Toda hierarchy in the dispersionless
limit (2dToda) or Sdiff(2) hierarchy \cite{TT}. Due to commutativity
of the 2dToda vector fields, the maps (\ref{transform}),
(\ref{transform1}) are functions of the deformation parameters
$\tau_1, \tau_2, ....$ $ \bar \tau_1, \bar \tau_2, ...$ (the 2dToda
``times"). The 2dToda system is an integrable Hamiltonian system of
PDEs for the coefficients $r(t,\tau_1,...,\bar \tau_1 ...)$,
$u_k(t,\tau_1,...,\bar \tau_1 ...)$, $\bar u_k(t,\tau_1,...,\bar
\tau_1 ...), k>0$ of the series (\ref{transform}),
(\ref{transform1}), obtained by equating (\ref{toda}) as Laurent
series in the dummy variable $w$. The string equation (\ref{string})
defines a reduction of the 2dToda hierarchy. The connection between the 2dToda
hierarchy and the Hele-Shaw problem was first found in \cite{MWZ},
\cite{WZ}.

In the context of the constant-coefficient Hele-Shaw problem, the
2dToda ``times" $\tau_k$ can be naturally interpreted as harmonic
moments of the domain $\Omega$. More precisely, the $k$th harmonic
moment $\int_{\Omega(t,\tau_1,...,\bar \tau_1,...)}z^kdXdY$ evolves
linearly in $\tau_k$ and is constant along the other 2dToda flows
\cite{HLY}, \cite{MWZ}, \cite{WZ}.

\end{section}

\begin{section}{Gauge Non-trivial Problems, Quantum Integrable Systems on Plane}

In the gauge trivial cases, equation (\ref{linearevolution}) can be
transformed into the differential ``string" equation (\ref{string})
for time dependent conformal maps. In contrast to the gauge-trivial
case, similar differential representations of the gauge-non-trivial
integral relations (\ref{linearevolution}) seem to be generically
impossible. Nevertheless, we can still construct explicitly the
quadrature domains for special gauge-non-trivial elliptic PDEs. In
this section, we derive sets of solutions to (\ref{schrodinger}) and
related conserved quantities that will be used for construction of
such quadrature domains.

An explicit evaluation of conserved quantities is possible when a
gauge-non-trivial elliptic operator $L$ in (\ref{def}) is equal, up
to a gauge transformation, (or (\ref{operator}) is equal) to a
Hamiltonian of a quantum integrable system on the plane.

We consider problems (\ref{norm}) with such $\kappa\eta$ that the
corresponding second-order elliptic differential operator $L$
(\ref{def}) can be related to the Laplace operator by a differential
operator $T$
\begin{equation}
T\Delta=LT \label{intertwinning}
\end{equation}
$T$ is usually called the intertwining operator and
(\ref{intertwinning}) is an intertwining identity. The simplest
intertwining identity corresponds to the gauge trivial case
(\ref{def}) when $T=T_0:=h^{-1}$ is a zero-order differential
operator and $L={\mathcal L}_0:=\frac{1}{h} \Delta h$. If, however,
$T$ is a differential operator of a non-zero order, the
corresponding $L$ equals, up to a gauge transformation, a
Hamiltonian of a non-trivial integrable quantum system on the plane.
Indeed, from (\ref{intertwinning}) it follows that any eigenfuction
$\Psi(z,\bar z,\lambda)$ of $\Delta$
$$
\Delta\left[\Psi(z,\bar z, \lambda)\right]=\lambda\Psi(z,\bar z,
\lambda)
$$
is transformed by the action of $T$, $\Psi(z,\bar z,\lambda)\to
T[\Psi(z, \bar z,\lambda)]$, to an eigenfunction of $L$ having the
same eigenvalue $\lambda$ or to zero.

We start with a class of examples, in which $\kappa, \eta$ vary only
in one direction (say $X$-direction) and equal
\begin{equation}
\kappa=\frac{1}{X^{2n}}, \quad \eta=1, \quad n=0,1,2,... \label{2n}
\end{equation}
The corresponding elliptic operator (\ref{def})
\begin{equation}
L={\mathcal L}_n:=X^{2n}\frac{\partial}{\partial
X}\frac{1}{X^{2n}}\frac{\partial}{\partial
X}+\frac{\partial^2}{\partial Y^2} \label{cartezian}
\end{equation}
is amenable, by a gauge transformation, to the two-dimensional
Schrodinger operator
\begin{equation}
{\mathcal H}_n=X^n{\mathcal L}_nX^{-n}=\frac{\partial}{\partial
Y^2}+\frac{\partial^2}{\partial X^2}-\frac{n(n+1)}{X^2} \label{a2}
\end{equation}
which , when rewritten down in the polar coordinates, has the
following form
\begin{equation}
{\mathcal H}_n=\frac{\partial^2}{\partial
\rho^2}+\frac{1}{\rho}\frac{\partial}{\partial
\rho}+\frac{S_n}{\rho^2}, \quad S_n=\frac{\partial^2}{\partial
\theta^2}-\frac{n(n+1)}{\cos(\theta)^2}, \quad z=X+iY=\rho
e^{i\theta} \label{polars}
\end{equation}
The operator $S_n$ admits the following alternative factorisations
$$
\begin{array}{l}
S_n=\left(\frac{\partial}{\partial\theta}+n\tan(\theta)\right)
\left(\frac{\partial}{\partial\theta}-n\tan(\theta)\right)-n^2=\\
\qquad\qquad
\left(\frac{\partial}{\partial\theta}-(n+1)\tan(\theta)\right)
\left(\frac{\partial}{\partial\theta}+(n+1)\tan(\theta)\right)-(n+1)^2
\end{array}
$$
and therefore
$$
\left(\frac{\partial}{\partial\theta}+(n+1)\tan(\theta)\right)S_n
=S_{n+1}\left(\frac{\partial}{\partial\theta}+(n+1)\tan(\theta)\right)
$$
In view of (\ref{cartezian}), (\ref{polars}), this leads to the
intertwining identity
\begin{equation}
T_n\Delta={\mathcal L}_nT_n, \quad
T_n=X^n\left(X\frac{\partial}{\partial Y}-Y\frac{\partial}{\partial
X}+(n+1)\frac{Y}{X}\right)\cdots\left(X\frac{\partial}{\partial
Y}-Y\frac{\partial}{\partial X}+\frac{Y}{X}\right) \label{link}
\end{equation}
Note, that the intertwining operator is not unique. To see this on the
example of equation (\ref{cartezian}), we can use the translational
invariance of the latter along the $Y$-direction, shifting $Y$ in $T_n$ by a
constant $\lambda$, and so obtaining a linear combination of
operators
$$
T_n+\sum_{i=1}^n\lambda^i T^{(i)}_n
$$
$T_n^{(i)}$ as well as $T_n$, are all intertwining operators of the
$n$th order. They are homogenous polynomials in
$X,Y,\frac{\partial}{\partial X},\frac{\partial}{\partial Y}$. For
instance, in the simplest $n=1$ case
\begin{equation}
T_1=X\left(X\frac{\partial}{\partial Y}-Y\frac{\partial}{\partial
X}\right)-Y, \quad T_1^{(1)}=X\frac{\partial}{\partial X}-1
\label{examplen=1}
\end{equation}
It is easy to see that operators $T^{(n)}_n$ (e.g. $T_1^{(1)}$ in
the above example) are $X$-dependent only, i.e.
\begin{equation}
T_n^{(n)}=X^n\frac{\partial^n}{\partial
X}+\sum_{k=0}^{n-1}a_{k;n}X^k\frac{\partial^k}{\partial X^k}
\label{help}
\end{equation}
Operator (\ref{help}) can be also obtained alternatively by exploiting
separation of (\ref{cartezian}) in Cartezian coordinates $X,Y$,
through a chain of factorizations leading from
$\frac{\partial^2}{\partial X^2}$ to $X^{
2n}\frac{\partial}{\partial
X}\frac{1}{X^{2n}}\frac{\partial}{\partial X}$.

Images of all intertwining operators, acting on harmonic functions,
coincide for a given $n$. Therefore, one can choose any operator
from $T_n, T^{(i)}_n, i=1..n$ (or their linear combination) to
construct
 same set of solutions to (\ref{cartezian}). In so doing, we return to
the $n=1$ example (\ref{examplen=1}), choosing $T^{(1)}_1$. In this
example, it is convenient to make a shift of $z$ by the distance
$x_1$ along the $X$ direction, displacing the singular line $X=0$ of
the problem (\ref{cartezian}) to $X=-x_1$. Then, without loss of
generality we can locate a source at point $z=0$. It follows that
the functions
$$
\phi_{k;1}=2T_1^{(1)}[z^{k+1}]=(k+1)\left(z+\bar
z+2x_1\right)z^k-2z^{k+1}, \quad k=0,1,2,...
$$
and their complex conjugates, form a set of solutions of equation
$$
\left((X+x_1)^2\frac{\partial}{\partial
X}\frac{1}{(X+x_1)^2}\frac{\partial}{\partial
X}+\frac{\partial^2}{\partial Y^2}\right)[\phi_k;1]=0
$$
related to the variable-coefficient Hele-Shaw problem with
$\kappa=1/(X+x_1)^2, \quad \eta=1$. Plugging this set into
(\ref{integrals}), by (\ref{linear}) we get an infinite number of
quantities
\begin{equation}
M_{k;1}=M[\phi_{k;1}], \label{examplen=1mom}
\end{equation}
that are linear and constant in time. They form a complete set of
local coordinates. A simple way to see the latter is to tend $x_1$
to infinity
$$
\frac{\phi_{k;1}}{2(k+1)x_1}=z^k+\frac{(k+1)(z+\bar
z)z^k-2z^{k+1}}{2(k+1)x_1}\to z^k, \quad {\rm as} \quad x_1
\to\infty
$$
observing that the set $\phi_{k;1}$ tends continuously to a basis
$z^k, k=0,1,..$ of functions analytic in a neighborhood of $z=0$.
The linearizing coordinates (\ref{examplen=1mom}) are, therefore in
one to one correspondence with harmonic moments
$\int_{\Omega}z^kdXdY$, at least in some neighborhood of infinity.
The corresponding variable-coefficient Hele-Shaw problem also
transforms continuously into the constant coefficient one. Since
harmonic moments are local coordinates for a generic set of
simply-connected domains $\Omega$ (see e.g \cite{VE} and references
therein), so are $M[\phi_k], k=0,1,...$.

A similar argument about completeness may be applied to the rest of
examples considered below. For instance, it is convenient to use the
following set of regular solutions to (\ref{cartezian}) when dealing
with an arbitrary $n$ problem for a flow, driven by a finite
combination of multipole sources located at $z=z_1$
\begin{equation}
\phi^{(n)}_0=1, \quad \phi^{(n)}_k(z, \bar z)=T_n^{(n)}[(z-z_1)^{n+k}], \quad \bar \phi^{(n)}_k(\bar z, z)=T_n^{(n)}[(\bar z-\bar z_1)^{n+k}], \quad
k\ge 1 \label{N}
\end{equation}

The following class of variable coefficient Hele-Shaw problem, whose
conserved quantities write in terms of polynomials in $\bar z, z$,
generalizes (\ref{2n})
\begin{equation}
\kappa=\frac{1}{(z^m+\bar z^m)^{2n}(z^m-\bar z^m)^{2l}}, \quad
\eta=1, \quad n>l\ge 0 \label{carusel}
\end{equation}
An elliptic operator (\ref{schrodinger}), corresponding to
(\ref{carusel})
\begin{equation}
L={\mathcal L}_{n,l;m}:=2\frac{\partial^2}{\partial z\partial \bar
z}-\frac{2nm}{z^m+\bar z^m}\left(z^{m-1}\frac{\partial}{\partial
\bar z}+\bar z^{m-1}\frac{\partial}{\partial
z}\right)-\frac{2lm}{z^m-\bar
z^m}\left(z^{m-1}\frac{\partial}{\partial \bar z}-\bar
z^{m-1}\frac{\partial}{\partial z}\right) \label{Calogero}
\end{equation}
equals, up to a gauge transformation, the Schrodinger operator of
the Calogero-Moser system related to (dihedral) group of symmetries
of a regular $4m$-polygon ($2m$-polygon if $l=0$).
$$
\begin{array}{c}
{\mathcal H}_{n,l;m}=(z^m+\bar z^m)^{n}(z^m-\bar z^m)^{l}{\mathcal
L}_{n,l;m}(z^m+\bar z^m)^{-n}(z^m-\bar
z^m)^{-l}=\frac{\partial^2}{\partial
\rho^2}+\frac{1}{r}\frac{\partial}{\partial \rho}+\frac{S_{n,l;m}}{\rho^2}, \\
S_{n,l;m}=\frac{\partial^2}{\partial
\theta^2}-\frac{m^2n(n+1)}{\cos(m\theta)^2}-\frac{m^2l(l+1)}{\sin(m\theta)^2}
\end{array}
$$
By analogy with $S_n$ in (\ref{polars}), $S_{n,l;m}$ admits
factorizations leading to the intertwining identity
$$
T_{n,l;m}\Delta={\mathcal L}_{n,l;m}T_{n,l;m}
$$
where the intertwining operator $T_{n,l;m}$ can be represented in a
form of the Wronskian \cite{BL}
\begin{equation}
T_{n,l;m}[f]=\rho^{m(n+l)}\frac{W[\psi_1,\psi_2,...,\psi_n,f]}{\cos(m\theta)^{n(n-1)/2}\sin(m\theta)^{l(l-1)/2}},
\quad W[f_1,..,f_k]:=\det\left[\frac{\partial^{j-1} f_i}{\partial
\theta^{j-1}}\right]_{1\le i,j\le k} \label{laplacetocalogero}
\end{equation}
with
$$\psi_k=
\left\{\begin{array}{l}
\sin\left(k(m\theta+\pi/2)\right), \quad k=1,2,..n-l \\
\cos\left((2k+l-n)(m\theta+\pi/2)\right), \quad k=n-l+1, n-l+2,..n
\end{array}
\right.
$$
$T_{n,l;m}$ is a differential operator of the $n$th order and is a
homogenous polynomial in $z, \bar z, \frac{\partial}{\partial z},
\frac{\partial}{\partial \bar z}$. It transforms any holomorphic
function $f(z)$ into a polynomial in $\bar z$, which is crucial for
construction of quadrature domains.

It is noteworthy that the intertwining operator $T_{n,l;m}$ is also
not unique. Although, in difference from the $l=0,m=1$ case we
cannot apply a simple argument connected with the translational
invariance of the system along the $Y$-direction, another
intertwining operator, constructed by the Dunkl method \cite{D},
exists. Various operators connecting $\Delta$ with ${\mathcal
L}_{n,l;m}$, that are compositions of the intermediate intertwining
operators constructed by the Dunkl method and those constructed
through factorizations, can be also obtained. Similarly to the
$l=0,m=1$ case, images of all so obtained intertwining operators,
acting on harmonic functions, coincide for fixed $n,l$, and to derive
a complete set of solutions to (\ref{Calogero}) one may
use just one of them, say (\ref{laplacetocalogero}).

\end{section}

\begin{section}{Construction of Gauge-non-trivial Quadrature Domains}

Let us start from gauge-non-trivial examples of the Hele-Shaw flows
(\ref{norm}), (\ref{2n}), that are driven by a multipole source
located at $z=z_1$. Since the quadrature domain $\Omega$ can be
viewed as a solution to the Hele-Shaw problem that develops
continuously in time, starting from zero initial data, its form is
defined by the condition that the quadrature identity
(\ref{identity1}) holds for any solution of (\ref{cartezian}),
regular in $\Omega$.

We show first, that curves, parametrized by polynomial maps of any
non-negative degree $ s \ge 0$
\begin{equation}
z(w)=z_1+rw+\sum_{k=2}^su_{k-1}w^{k}, \quad z(e^{i\theta}) \in
\partial \Omega,
\label{transform11}
\end{equation}
where $r$ can be made to be real, are boundaries of the quadrature
domains for solutions of (\ref{cartezian}). As in the
constant-coefficient problem, forms, sizes and positions of these
domains are functions of $s+1$ free complex parameters
$z_1,r,u_k,k=2..s$.

As follows from the previous section, any solution to
(\ref{cartezian}), that is regular in $\Omega$ can be represented as
(\ref{help})
$$
\phi=2T^{(n)}_n[f(z)]=\sum_{k=0}^n\frac{a_{k;n}}{2^k}(z+\bar
z)^k\frac{\partial^k f(z)}{\partial z^k}, \quad a_{n;n}=1
$$
where $f(z)$ is analytic in vicinity of $z=z_1$. Substituting this
solution into (\ref{identity1}) and using the Green theorem, we can
transform the integral over $\Omega$ in the left-hand side of
(\ref{identity1} ) to the line integral
$$
\int_{\Omega}\phi
dXdY=\frac{1}{2i}\oint_{|w|=1}\sum_{k=0}^n\frac{a_{k;n}}{2^k(k+1)}\left(\left(z(w)+\bar
z(1/w)\right)^{k+1}\left(\frac{1}{\frac{\partial z}{\partial
w}}\frac{\partial }{\partial w}\right)^k[f(z(w))]\right)dw
$$
Using (\ref{transform11}) and taking analyticity of $f(z)$ into
account, we get
\begin{equation}
\int_{\Omega}\phi
dXdY=\sum_{i=0}^{(n+1)(s+1)-2}U_i(z_1,r,u_1,..,u_s,\bar u_1,..,\bar
u_s)\left(\frac{\partial^i f(z)}{\partial z^i}\right)_{z=z_1}
\label{eq1}
\end{equation}
Comparing (\ref{eq1}) with the right-hand side of (\ref{identity1}),
we see that the latter must contain $s(n+1)-1$ derivatives of $\phi$
and is equal to
\begin{equation}
\sum_{i=0}^{(s+1)(n+1)-2}V_i\left(\frac{\partial^i f(z)}{\partial
z^i}\right)_{z=z_1} \label{eq2}
\end{equation}
where, $V_i$ are linear functions of
$Q_1^{(0)},..Q^{(s(n+1)-1)}_1,\bar Q^{(0)}_1,..\bar
Q^{(s(n+1)-1)}_1$. Eqs. (\ref{eq1}) and (\ref{eq2}) lead to the
non-homogenous over-determined linear system of $2(s+1)(n+1)-1$
equations
\begin{equation}
\quad U_i=V_i, \quad \bar U_i=\bar V_i, \quad i=0..(s+1)(n+1)-2,
\quad Q^{(0)}_1=\bar Q^{(0)}_1 \label{system}
\end{equation}
for $2s(n+1)$ unknowns $Q_1^{(k)},\bar Q^{(k)}_1,k=0..s(n+1)-1$.
Although the number of equations exceeds the number of unknowns by
$2n+1$, not all these equations are independent. For instance,
solving (\ref{system}) for circular domains, $s=1, z=z_1+rw,$ we get
the simplest gauge-non-trivial quadrature identities
$$
\int_{|z-z_1|\le r}\phi dXdY=\pi r^2\phi(z_1)+\frac{\pi
r^4}{4x_1}\left(\frac{\partial \phi}{\partial z}+\frac{\partial
\phi}{\partial \bar z}\right)_{z=z_1}, \quad z_1=x_1+iy_1
$$
$$
\int_{|z-z_1|\le r}\phi dXdY=\pi r^2\phi(z_1)+\frac{\pi
r^4(r^2+12x_1^2)}{24x_1^3}\left(\frac{\partial \phi}{\partial
z}+\frac{\partial \phi}{\partial \bar z}\right)_{z=z_1}+\frac{\pi
r^6}{24x_1^2}\left(\frac{\partial^2 \phi}{\partial
z^2}+\frac{\partial^2 \phi}{\partial \bar z^2}\right)_{z=z_1}
$$
for solutions of (\ref{cartezian}) with $n=1$ and $n=2$
respectively. Moreover, a circular solution to the arbitrary $n$
variable-coefficient Hele-Shaw problem (\ref{2n}) exists, if the
flow is driven by a special combination of a monopole, dipole ...
$n+1$-pole sources located at the same point $z_1$.

To show that the polynomial solutions of an arbitrary non-negative
degree exist for any $n$ in (\ref{2n}), (i.e that the number of
independent equations in arbitrary $n,s$ case equals the number of
unknowns $Q_1^{(i)},\bar Q^{(i)}_1$ and these equations are
compatible), it is more convenient to check the quadrature identity
(\ref{identity1}) on a complete set (\ref{N}) of solutions to
(\ref{cartezian}).

According to (\ref{eq1}), (\ref{eq2}) and (\ref{N}), we have to
evaluate
\begin{equation}
\sum_{j=0}^{(s+1)(n+1)-2}(V_j-U_j)\frac{\partial^jf(z)}{\partial z},
\quad \sum_{j=0}^{(s+1)(n+1)-2}\left(\bar V_j-\bar
U_j\right)\frac{\partial^jf(\bar z)}{\partial \bar z^j}
\label{argumentation}
\end{equation}
on $f(z)=(z-z_1)^{n+k}$ (and $f(\bar z)=(\bar z-\bar z_1)^{n+k}$
respectively) at point $z=z_1$ for $k\ge 1$. Also there is one more
equation, obtained by substituting $\phi=\phi_{0;1}^{(n)}=1$ in
(\ref{identity1}). The latter is equation for $Q_1^{(0)}$, that, as
seen from (\ref{identity1}), has a real solution, and equation
$Q_1^{(0)}=\bar Q_1^{(0)}$ in (\ref{system}) is satisfied. So, there
remains $2s(n+1)-2$ unknowns $Q^{(j)}_1, \bar Q^{(j)}_1,
j=1..s(n+1)-1$.

The highest derivatives in (\ref{argumentation}) are of the order
$(s+1)(n+1)-2$, so that $(z-z_1)^j$ (or $(\bar z-\bar z_1)^j$ for
the second equation in (\ref{argumentation})) is annihilated at
$z=z_1$ if $j>(s+1)(n+1)-2$. As a consequence, (\ref{argumentation})
is identically satisfied for $k>s(n+1)-1$ in (\ref{N}), and there
remains $2s(n+1)-2$ equations and the equal number of unknowns
$Q^{(j)}_1, \bar Q_1^{(j)}, j=1..s(n+1)-1$.

We now have to prove the compatibility of remaining equations.
(\ref{argumentation}) is a non-homogenous system of linear equations
for unknowns $Q_1^{(0)},Q^{(k)}_1,\bar Q^{(k)}_1$, fixing their
dependence on parameters $z_1,\bar z_1,r,u_1,...,u_s,$ $\bar
u_1,...,\bar u_s$. Such a system is compatible if its determinant
does not vanish i.e. its homogenous part does not have nontrivial
solutions. Let us suppose that it does. Remind that the homogenous
part of the system has been obtained by the action of the operator
\begin{equation}
\hat
Q_1=Q^{(0)}_1+\sum_{j=1}^{s(n+1)-1}\left(Q^{(i)}_1\frac{\partial^i}{\partial
z^i}+\bar Q_1^{(i)}\frac{\partial^i}{\partial \bar z^i}\right)
\label{ator}
\end{equation}
in the right hand-side of (\ref{identity1}), to an arbitrary
solution of (\ref{cartezian}) at point $z=z_1$. So, if the
determinant of the system vanished, then there would exist such
$Q^{(0)}_1, Q^{(k)}_1, \bar Q^{(k)}_1, k=1..(s+1)n-1$, that the
operator (\ref{ator}) would annihilate any solution of
(\ref{cartezian}) at $z=z_1$, i.e.
\begin{equation}
\hat Q_1 T^{(n)}_n[f(z)]=0, \quad {\rm at} \quad z=z_1
\label{action}
\end{equation}
where $f(z)$ is any analytic in $\Omega$ function. If the above were
true, then changing continuously the position $z=z_1$ of the source,
we could pick such coefficients $Q^{(0)}_1=Q^{(0)}_1(z,\bar z),
Q^{(i)}_1=Q^{(i)}_1(z,\bar z), \bar Q^{(i)}_1=\bar Q^{(i)}_1(z,\bar
z)$, that (\ref{action}) would hold at each point in some region of
the plane. In so doing we could construct a such operator $\hat
Q_1$, with coefficients depending on $z$, that $\hat
Q_1T_n^{(n)}[f(z)]=0$ in some region of the plane for an arbitrary
$f(z)$. But this is evidently impossible, since the highest symbols
of $\hat Q_1$ and $T_n^{(n)}$ contain pure derivatives in $z$, so
does their composition.

Thus, the system of equations (\ref{system}) is compatible and has a
unique solution
$$
Q_1^{(0)}=Q_1^{(0)}(r,u_1,..,u_s,\bar u_1,..\bar u_s), \quad
Q_1^{(k)}=Q_1^{(k)}(x_1,r,u_1,..,u_s, \bar u_1,..\bar u_s), \quad
k=1..s(n+1)-1
$$
defining quadrature domains with boundaries parametrized by
polynomial conformal maps (\ref{transform11}) of an arbitrary
non-negative degree $s\ge 0$. These domains are solutions to the
Hele-Shaw problems, describing the free-boundary flows driven by a
multipole (combination of a monopole, dipole, ...., $s(n+1)$-pole)
source located at point $z=z_1$.

Analogous analysis can be applied to the problems related to general
dihedral systems (\ref{carusel}), (\ref{Calogero}). It also leads to
the conclusion about existence of polynomial solutions
(\ref{transform11}) that depend on $s+1$ free parameters. The flow,
in this case, is driven by a combination of a monopole, dipole,
..$s(m(n+l)+1)$-pole sources located at $z=z_1$.

\end{section}

\begin{section}{Conclusion}

We have shown that, similarly to the constant coefficient Hele-Shaw
problem, a class of the variable coefficient problems, connected
with the dihedral quantum Calogero-Moser systems, admits polynomial
solutions for the flows driven by a finite number of multipole
sources. In particular, the spaces of polynomial solutions have same
dimensions in both the constant and variable coefficient cases.

\end{section}

\vspace{5mm}

{\bf \Large Acknowledgement}\\

The author would like to acknowledge useful information and help
received from P. Etingof, S. Howison and J. Ockendon. This work is
supported by the European Community IIF MIF1-CT-2005-007323.


\vspace{5mm}

\begin{thebibliography} {Bibliography}

\bibitem {BL} Yuri Yu. Berest, Igor M. Loutsenko, Huygens' principle in Minkowski spaces and soliton solutions of the Korteveg-de Vries Equation, Comm.Math.Phys. 190, 113-132 (1997), solv-int/9704012

\bibitem{D} Dunkl C.F., Differential-difference operators associated
to reflection groups, Trans.Amer.Math.Soc. 311 (1989), 181-191

\bibitem{EE} Entov, V. M., Etingof, P. I. Bubble contraction in Hele-Shaw cells. Quart.
J. Mech. Appl. Math.  44  (1991),  no. 4, 507--535

\bibitem {G} Galin, L. A. Unsteady filtration with a free surface. C. R.
(Doklady) Acad. Sci. URSS (N.S.) 47, (1945). 246--249

\bibitem {HLY} J. Harnad, I. Loutsenko, O. Yermolayeva, Constrained
reductions of the 2dToda hierarchy, Hamiltonian structure and
interface dynamics, to appear in J.Math.Phys, math-ph/0312058

\bibitem{H} Howison, S. D. , Fingering in Hele-Shaw cells, J.Fluid Mech, 167 (1986)

\bibitem{MD}  Mark B. Mineev-Weinstein, Silvina Ponce Dawson, A New Class of Nonsingular Exact Solutions for Laplacian Pattern Formation, Phys. Rev. E 50, R24 (1994)
patt-sol/9305010

\bibitem{MWZ}  M. Mineev-Weinstein, P. Wiegmann, A. Zabrodin, Integrable
Structure of Interface Dynamics, Phys. Rev. Lett. 84,
(2000)5106-5109, nlin.SI/0001007

\bibitem {P} Polubarinova-Kotschina, P. J. On the displacement of the oil-bearing
contour. C. R. (Doklady) Acad. Sci. URSS (N. S.) 47, (1945).
250--254

\bibitem {R} Richardson S. Hele Shaw flows with a free boundary produced
by the injection of fluid into a narrow channel. J.Fluid Mech., 56,
part 4, (1972), pp.609-618

\bibitem{Sa} Sakai M. Quadrature domains, Lecture Notes in Mathematics, (1978) 934, Springer-Verlag

\bibitem{Sh} H.S. Shapiro, The Schwartz function and its generalization
to higher dimension, Wiley, new York, (1992)

\bibitem{TT} K.Takasaki, T.Takebe, Integrable hierarchies and dispersionless limit. Rev. Math. Phys. 7 (1995), no. 5, 743--808, hep-th/9405096

\bibitem{VE} A.N. Varchenko and P.I. Etingof, Why the boundary of a round drop becomes a curve of order four, American Mathematical Society, University Lecture Series, 3, (1994)

\bibitem{WZ} Wiegmann, P. B.; Zabrodin, A., Conformal maps and integrable hierarchies, Comm. Math. Phys. 213 (2000), no. 3, 523--538, hep-th/9909147

\end{thebibliography}
\end{document}